\documentclass[11pt]{article}
\usepackage{dsfont, amssymb,amsmath,amscd,latexsym, amsthm, amsxtra,amsfonts, mathrsfs}
\usepackage{graphicx}
\usepackage[all]{xy}
\begin{document}
\baselineskip 18pt
\title{{ {Approximating Partition Functions of Two-State Spin Systems}}}
\author{Jinshan \ Zhang\thanks{Corresponding author:
zjs02@mails.tsinghua.edu.cn} Heng Liang and Fengshan Bai
\\ \small Department of Mathematical Sciences,
\  Tsinghua University,\\
\small Beijing 100084,  China }
\date{}
\maketitle

{\small\begin{center}\textbf{Abstract}\end{center}

Two-state spin systems is a classical topic in statistical physics.
We consider the problem of computing the partition function of the
systems on a bounded degree graph. Based on the self-avoiding tree,
we prove the systems exhibits strong correlation decay  under the
condition that the absolute value of  ``inverse temperature" is
small. Due to strong correlation decay property,  an FPTAS for the
partition function is presented under the same condition. This
condition is sharp for Ising model.\\\\
\text{\\ \textsl{Keywords:} Strong correlation decay; Self-avoiding
tree; Ising model; FPTAS; Partition function}}\\\\\\
\textbf{\large{ 1. Introduction}}

Spin model with $p$ states is a classical mathematical model in
statistical physics. Such models describe and explain the behavior
of ferromagnets, lattice gas and certain other phenomena of
statistical physics. In this paper, we focus on the case of two
spins. This case encompasses models of physical interest, such as
the classical Ising model (ferromagnetic or antiferromagnetic, with
or without an applied magnetic field).

In statistical mechanics, the partition function is an important
quantity that encodes the statistical properties of a system in
thermodynamic equilibrium. However, partition functions are normally
hard to compute, even for the two-state spin systems \cite{GJP03}.
Markov Chain Monte Carlo methods \cite{JS93,MS08} are the existing
powerful approach. 

Exploiting the structure property of Gibbs measure, Weitz
\cite{We06} and Bandyopadhyay, Gamarnik \cite{BG06} introduce new
deterministic algorithm for counting the number of independent sets
and colorings. The key point of this method is to establish the
$strong$ $spatial$ $mixing$ property, which is also known as
$strong$ $correlation$ $decay$, on certain defined rooted  trees. It
follows that the marginal probability of the root is asymptotically
independent of the configuration on the leaves far below. In
\cite{We06}, Weitz proves the strong correlation decay for hard-core
model on bounded degree trees and pushes the result to general graph
using the $self$-$avoiding$ $tree$ technique. The proof employs the
recursive formula for computing the marginal probability of a vertex
on the tree. This approach  is well known for some kinds of
statistical systems, such as Ising model \cite{PP06} and coloring
model \cite{Jo02}.


It is natural to ask whether the more general two-state spin systems
exhibits strong correlation decay. We present a positive answer
based on the recursive formula on bounded trees  in this paper.  We
show that, for arbitrary external field, the Gibbs measure exhibits
strong correlation decay on a bounded degree tree when the absolute
value of the inverse temperature is smaller  than $J_d$, where $J_d$
is critical point for uniqueness of Gibbs measures of
(anti)ferromagnetic Ising model on an infinite $d$ regular
tree\cite{Ge88,We05}. This generalizes the recent result by Mossel
and Sly\cite{MS08}. They prove the strong correlation decay for
ferromagnetic Ising model. By the strong correlation decay, we prove
that there exists an unique Gibbs measure of two-state spin systems
on an infinite bounded degree graph. This generalizes the
Dobrushion's condition, $d\tanh(J)<1$ to $(d-1)\tanh(J)<1$, for the
uniqueness of Gibbs measure of antiferromagnetic and ferromagnetic
Ising models \cite{Do70,Ge88,We05}. Since an infinite $d$ regular
tree is a special infinite bounded degree graph, the condition is
sharp for Ising model.

A fully polynomial time approximation schemes (FPTAS) for partition
functions of two-state spin systems on a bounded degree graph is
presented, which is natural  and reasonable when the strong
correlation decay holds. Jerrum and Sinclair \cite{JS93} provided an
FPRAS to ferromagnetic Ising model for graphs with any uniform
positive inverse temperature and identical external field for all
the vertices. Their results do not include the case where different
vertices have different external field, and are not applied to
antiferromagnetic Ising model either. Very recently Dembo and
Montanari propose an explicit formula for partition function of
ferromagnetic Ising model with any external field on locally
tree-like graphs, which still does not include the antiferromagnetic
case\cite{DM08}.

The remainder of the paper has the following structure. In Section
2, we present some preliminary definitions . We go on to prove the
main theorem in Section 3. Section 4 is devoted to propose an FPTAS
for the partition functions under our conditions. Further work and
conclusion are given in Section 5.\\\\
\textbf{\large{2. Notations and Definitions}}

Let $G=(V,E)$ be a finite graph with vertex set $V=\{1,2,\cdots,n\}$
and edge set $E$. Let $d(u,v)$ denote the distance between $u$ and
$v$, for any $u$,$v\in V$. A path $v_1\rightarrow v_2\rightarrow
\cdots$ is called a self-avoiding path if $v_i\neq v_j$ for all
$i\neq j$. The distance between a vertex $v\in V$ and a subset
$\Lambda\subset V$ is defined by
$$d(v,\Lambda)=\min\{d(v,u):u\in\Lambda\}.$$ The set of vertices with
distance  $l$ to the vertex $v$ is denoted by $$S(G,v,l)=\{u:
d(v,u)= l\}.$$
The set of vertices which are no more than $l$ away from $v$ is
denoted by
$$V(G,v,l)=\{u: d(v,u)\leq l\}.$$
Let $\delta_v$ denote the degree of $v$ in $G$ and
$\Delta(G)=\max\{\delta_v: v\in V\}$. Let all the vertices in graph
$G=(V,E)$ be numbered, where $V$ and $E$ are vertex set and edge set
of $G$ respectively. We define the partial order on $E$, where
$(i,j)>(k,l)$ if and only if $(i,j)$ and $(k,l)$ share a common
vertex and $i+j > k+l$. In two-state spin systems on $G$, each
vertex $i\in V$ is associated with a random variable $X_i$ on
$\Omega=\{\pm1\}$ ($\pm$ in brief).\\\\
 {\textbf{Definition 1.}} The
Gibbs measure of two-state spin systems on $G$ is defined by the
joint distribution of $X=\{X_1,X_2,\cdots,X_{n}\}$
\begin{displaymath} P_G(X=\sigma)=\frac{1}{Z(G)}\exp(\sum\limits_{(i,j)\in
E}\beta_{ij}(\sigma_i,\sigma_j)+\sum\limits_{i\in V}h_i(\sigma_i)).
\end{displaymath}
where $h_i$ is a map $\Omega\rightarrow R$ and $\beta_{ij}$ is a map
$\Omega^2\rightarrow R$. $Z(G)$ is called the partition function of
the system.

Note that the Gibbs measure would satisfy  
$\sum_{\sigma\in\Omega^n}P_G(X=\sigma)=1$. We use notation
$\beta_{ij}(a,b)=\beta_{ji}(b,a)$. For any $\Lambda\subseteq V$,
$\sigma_{\Lambda}$ denotes the set $\{ \sigma_i, i\in \Lambda\}$.
With a little abuse of notation, $\sigma_{\Lambda}$ also denotes the
condition or configuration $\sigma_i$ with fixed $i$ for any $
i\in\Lambda$. Let $Z(G,\Phi)$ denote the partition function under
the condition $\Phi$, e.g. $Z(G,X_1=+)$ represents the partition
function under the condition the vertex $1$ is fixed $+$.\\
\begin{figure}[ht]
\centering
\includegraphics[scale=.4]{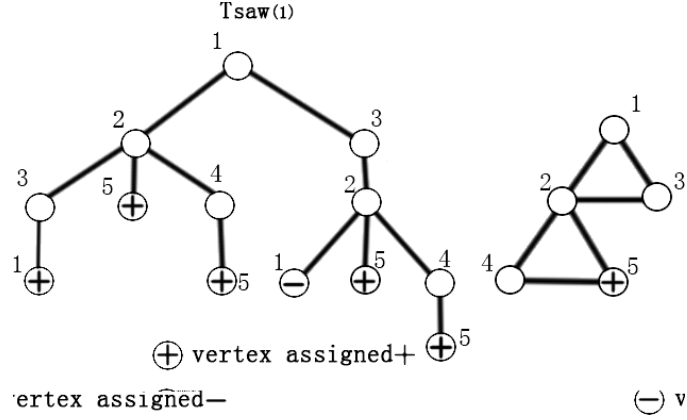}
\caption{\small  The graph with one vertex assigned + (Right)  and
its corresponding self-avoiding tree $T_{saw(1)}$ (Left)}
\end{figure}

A self-avoiding walk (SAW) is a sequence of moves (on a graph) which
does not visit the same point more than once. The following gives an
important tool in proving our results. It is
introduced in\cite{We06}.\\\\
\noindent \textbf{Definition 2.} (Self-Avoiding Tree) The
self-avoiding tree $T_{saw(v)}(G)$ (for simplicity denoted by
$T_{saw(v)}$) corresponding to the vertex $v$ of $G$ is the tree
with root $v$ and generated through the self-avoiding walks
originating at $v$. A vertex closing a cycle is included as a leaf
of the tree and is assigned to be $+$, if the edge ending the cycle
is larger
than the edge starting the cycle, and $-$ otherwise.\\

\noindent \emph{Remark:}  Given any configuration $\sigma_{\Lambda}$
of $G$, $\Lambda\subset V$, the self-avoiding tree is constructed in
the same way as the above procedure except that the vertex, which is
a copy of the vertex $i$ in $\Lambda$, is fixed to the same spin
$\sigma_i$ as $i$ and the subtree below it is not constructed due to
the Markov property, see Figure 1 for example, where vertex $5$ is
fixed $+$ in $G$. 
\\\\
 \textbf{Definition 3.} (Strong Correlation Decay)  The
Gibbs distribution of two-state spin systems on $G$ exhibits strong
correlation decay if and only if for any vertex $v\in V$, subset
$\Lambda\subset V$, any two configurations $\sigma_{\Lambda}$ and
$\eta_{\Lambda}$ on $\Lambda$, denote $t=d(v,\Theta)$, where
$\Theta=\{v\in\Lambda: \sigma_v\neq \eta_v\}$, there exits positive
numbers $a$, $b$ independent of $n$ such that
\begin{displaymath}
|\log P_G(X_v=+|\sigma_{\Lambda})-\log
P_G(X_v=+|\eta_{\Lambda})|\leq f(t),\end{displaymath} where decay
function $f(t)= a\exp(-bt)$.\\\\
\textbf{Definition 4.} (FPTAS) An approximation algorithm is called
a fully polynomial time approximation scheme (FPTAS) if and only if
for any $\epsilon>0$, it takes a polynomial time of input and
$\epsilon^{-1}$ to output a value $\bar{M}$ satisfying
\begin{displaymath}
e^{-\epsilon}\leq\frac{\bar{M}}{M}\leq e^{\epsilon},
\end{displaymath}
where $M$ is the real value.\\\\
\textbf{\large{3. Strong Correlation Decay}}





In two-state spin systems, when
$\beta_{ij}(\sigma_i,\sigma_j)=J_{ij}\sigma_i\sigma_j$,
$h_i=B_i\sigma_i$ for all the edge $(i,j)\in E$ and vertex $i\in V$,
and $J_{ij}$ is uniformly negative$/$positive for all $(i,j)\in E$,
the system is called antiferromagnetic/ferromagnetic Ising model.
Let
$$J_{ij}=\frac{\beta_{ij}(+,+)+\beta_{ij}(-,-)-\beta_{ij}(-,+)-\beta_{ij}(+,-)}{4},$$
and $B_i=\frac{h_i(+)-h_i(-)}{2}$ for all edges and vertices. We
call $J_{ij}$ and $B_i$ `inverse temperature' and `external field'
of two-state spin systems. Let $J=\max_{(i,j)\in E}|J_{ij}|$.  The
main theorem in this
section is summarized as follows\\\\
\textbf{Theorem 1.} \emph{Let $G=(V,E)$ be a graph with vertex set
$V=\{1,2,\cdots,n\}$ and edge set $E$.  There exists a numbers $d>0$
such that $\Delta(G)\leq d$. Suppose
\begin{displaymath}
(d-1)\tanh{J}<1,
\end{displaymath}
that is equivalent to $J<J_d=\frac{1}{2}\log(\frac{d}{d-2})$, then
the Gibbs distribution of the two-state spin systems on $G$ exhibits
 strong correlation decay for arbitrary external field.
Specifically, decay function is $$f(t)=4Jd((d-1)\tanh J)^{t-1}.$$}

In order to prove Theorem 1, four technical lemmas are given first.
The inequality in Lemma 1 is inspired by a
similar result in \cite{Ly89}. \\\\
\textbf{Lemma 1.}\emph{ Let $a$, $b$, $c$, $d$, $x$, $y$ be positive
numbers, $g(x)=\frac{ax+b}{cx+d}$ and
$t=|\frac{\sqrt{ad}-\sqrt{bc}}{\sqrt{ad}+\sqrt{bc}}|$, then
\begin{displaymath}
\max(\frac{g(x)}{g(y)},
\frac{g(y)}{g(x)})\leq(\max(\frac{x}{y},\frac{y}{x}))^t.
\end{displaymath}}\\
\textbf{Proof.} We separate the proof into two cases.

\noindent \textbf{Case 1.} $ad\geq bc$. Consider a function $$g(x)=
\frac{ax+b}{cx+d}=\frac{a}{c}-\frac{ad-bc}{c(cx+d)}.$$ It is clearly
an increasing function. Without loss of the generality, suppose
$x\geq y$ and let $x=zy$, where $z\geq1$, then
\begin{displaymath}
\begin{split}
\log(\frac{g(x)}{g(y)})&=\int^{z}_{1}\frac{d(\log(\frac{g(\alpha
y)}{g(y)}))}{d\alpha}d\alpha =\int^{z}_{1} (\frac{ay}{a\alpha
y+b}-\frac{cy}{c\alpha
y+d})d\alpha\\
&=\int^{z}_{1} \frac{(ad-bc)y}{(a\alpha y+b)(c\alpha y+d)}d\alpha\\
&=\int^{z}_{1} \frac{(ad-bc)y}{(\sqrt{ac}\alpha y-\sqrt{bd})^2+(\sqrt{bc}+\sqrt{ad})^2\alpha y}d\alpha\\
&\leq\int^{z}_{1} \frac{(ad-bc)y}{(\sqrt{bc}+\sqrt{ad})^2\alpha
y}d\alpha =\frac{\sqrt{ad}-\sqrt{bc}}{\sqrt{ad}+\sqrt{bc}}\log z.
\end{split}
\end{displaymath}
Hence
\begin{displaymath}
\max(\frac{g(x)}{g(y)}, \frac{g(y)}{g(x)})=\frac{g(x)}{g(y)}\leq
(\frac{x}{y})^t= (\max(\frac{x}{y},\frac{y}{x}))^t,
\end{displaymath}
where $t=\frac{\sqrt{ad}-\sqrt{bc}}{\sqrt{ad}+\sqrt{bc}}$.\\
\textbf{Case 2.} $ad\leq bc$. Now $g(x)$ is a decreasing function.
Let $h(x)=1/g(x)$, then $h(x)$ is an increasing function. Suppose
$x\geq y$, it is easy to be obtained that
\begin{displaymath}
\frac{h(x)}{h(y)}\leq
(\frac{x}{y})^{\frac{\sqrt{bc}-\sqrt{ad}}{\sqrt{ad}+\sqrt{bc}}}.
\end{displaymath}
Hence
\begin{displaymath}
\max(\frac{g(x)}{g(y)},
\frac{g(y)}{g(x)})=\frac{g(y)}{g(x)}=\frac{h(x)}{h(y)}\leq
(\frac{x}{y})^t= (\max(\frac{x}{y},\frac{y}{x}))^t,
\end{displaymath}
where $t=\frac{\sqrt{bc}-\sqrt{ad}}{\sqrt{ad}+\sqrt{bc}}$. \ \ \
$\Box$\\\\
\textbf{Lemma 2.}\emph { Let $T=T(V,E)$ be a rooted tree with vertex
set $V=\{0,1,2,\cdots,n\}$ and edge set $E$. The root is vertex $0$.
Suppose some vertices are fixed(assigned to certain spins) on $T$.
Removing an edge $(k,l)$, where $d(k,0)<d(l,0)$, let $T_k$ and $T_l$
be two resulting subtrees of $T$ including vertex $k$ and $l$
respectively. The fixed vertices remain fixed on $T_k$ and $T_l$.
Then the probability $P(X_0=+\ on\ T)$ equals the probability
$P(X_0=+\ on\ the\ subtree\ T_k)$ except
changing the `external field' $h_k$ to certain value $h^{'}_k$ on $T_k$.}\\\\
\textbf{Proof.} Let $\Omega_{T_l}$ denote the configuration space on
$T_l$. $E_l$ and $V_l$ denote the edge set and vertex set on $T_l$.
The following equality implies the result of the lemma.
\begin{displaymath}
\begin{split}
h^{'}_k(\sigma_k)&=h_k(\sigma_k)+
\log(\sum\limits_{\tau\in\Omega_{T_l}}e^{
\beta_{kl}(\sigma_k,\tau_l)+ \sum\limits_{(i,j)\in
E_l}\beta_{ij}(\tau_i,\tau_j)+\sum\limits_{i\in V_l}h_i(\tau_i)}). \
\ \ \ \ \ \  \Box
\end{split}
\end{displaymath}

With Lemma 1 and Lemma 2,  strong
correlation decay property on trees will be proved.\\\\
\textbf{Lemma 3.} \emph { Let $T=T(V,E)$ be a rooted tree with
vertex set $V=\{0,1,2,\cdots,n\}$ and edge set $E$. The root is
vertex $0$. Consider the two-state spin systems on it. Let
$\Lambda\subset V$ , $\zeta_{\Lambda}$ and $\eta_{\Lambda}$ be any
two configurations on $\Lambda$. Let $\Theta=\{i: \zeta_i\neq
\eta_i, i\in\Lambda\}$, $t=d(0,\Theta)$ and $s=|S(T,0,t)|=|\{i:
 d(0,i)=t, i\in T\}|$. Then
\begin{displaymath}
\begin{split}
\max(\frac{P_T(X_0=+|\zeta_{\Lambda})}{P_T(X_0=+|\eta_{\Lambda})}
,\frac{P_T(X_0=+|\eta_{\Lambda})}{P_T(X_0=+|\zeta_{\Lambda})})\leq
e^{4Js(\tanh{J})^{t-1}}.
\end{split}
\end{displaymath}}\\
\textbf{Proof.} For any $i\in V$, let $T_i$ denote the subtree with
$i$ as its root and $Z(i)$  be the two-state spin systems induced on
$T_i$ by $T$. Note that $T_0$ is equal to $T$.  To prove the
theorem, it's convenient to deal with the ratio
$\frac{P_T(X_0=+|\zeta_{\Lambda})}{P_T(X_0=-|\zeta_{\Lambda})}$
rather than $P_T(X_0=+|\zeta_{\Lambda})$ itself. Denote
$R^{\zeta_{\Lambda}}_i\equiv
\frac{P_{T_i}(X_i=+|\zeta_{\Lambda_{i}})}{P_{T_i}(X_i=-|\zeta_{\Lambda_{i}})}$,
where $\zeta_{\Lambda_{i}}$ is the condition  by imposing the
configuration $\zeta_{\Lambda}$ on $T_i$. When $x_1, x_2 \in (0,1)$,
$\frac{x_1}{x_2}\geq 1$ if and only if $\frac{x_1}{1-x_1}\geq
\frac{x_2}{1-x_2}$. Then
$$\max\{\frac{x_1}{x_2},\frac{x_2}{x_1}\}\leq
\max\{\frac{x_1/(1-x_1)}{x_2/(1-x_2)},\frac{x_2/(1-x_2)}{x_1/(1-x_1)}\}.$$
Replace $x_1$ and $x_2$ by $P_T(X_0=+|\zeta_{\Lambda})$ and
$P_T(X_0=+|\eta_{\Lambda})$. Then lemma 2 follows by
$$\max(\frac{P_T(X_0=+|\zeta_{\Lambda})}{P_T(X_0=+|\eta_{\Lambda})}
,\frac{P_T(X_0=+|\eta_{\Lambda})}{P_T(X_0=+|\zeta_{\Lambda})}) \leq
\max(\frac{R^{\zeta_{\Lambda}}_0}{R^{\eta_{\Lambda}}_0}
,\frac{R^{\eta_{\Lambda}}_0}{R^{\zeta_{\Lambda}}_0}).$$ Hence what
is needed to be proved becomes
\begin{equation}
\max(\frac{R^{\zeta_{\Lambda}}_0}{R^{\eta_{\Lambda}}_0}
,\frac{R^{\eta_{\Lambda}}_0}{R^{\zeta_{\Lambda}}_0})\leq
\exp(4Js(\tanh{J})^{t-1}).
\end{equation}
The inequality (1) is proved by induction on $t$. Before doing it,
some trivial cases need to be clarified. We are interested in the
case $t\geq 1$ and $0$ is unfixed. Let $\Gamma_{kl}$ denote the
unique self-avoiding path from $k$ to $l$ on $T$. If $i$ is a leaf
on $T$ and $d(0,i)<t$, where $t=d(0,\Theta)$, define $U_i=\{j\in V:
j\in \Gamma_{0i}, \exists k\in S(T,0,t), s.t. j\in\Gamma_{0k}\}$.
$U_i\neq \emptyset$ because of $0\in U_i$. Let $j_i\in U_i$ such
that $d(i,j_i)=d(i,U_i)$. By lemma 2, we can remove the subtree
bellow $j_i$ and change external field from $h_{j_i}$ to
$h^{'}_{j_i}$ at $j_i$ without changing the probability $P(X_0=+)$.
It is noted that this procedure removes at least one leaf with the
hight $< t$, and does not remove any vertex with the hight $\geq t$.
We can suppose that $T$ is a tree rooted at $0$ and the height of
every leaf on the tree is no less than $t$. Let $0_1,0_2,\cdots,0_q$
be the neighbors connecting with $0$.  The recursive formula can be
presented. Let $\Omega_{T_i}$ denote the configuration space in
$T_i$ under the condition $\zeta_{\Lambda}$, $i=1,2,\cdots,q$ and
$\Omega_{0}$ denote the configuration space of $T_0$ under the
condition $\zeta_{\Lambda}\cup\{\sigma_0\}$. We have
\begin{displaymath}
\begin{split}
&R^{\zeta_{\Lambda}}_0=\frac{Z(T_0,X_0=+,\zeta_{\Lambda})}{Z(T_0,X_0=-,\zeta_{\Lambda})}\\
&=\frac{e^{h_0(+)}\sum\limits_{\sigma\in\Omega_0}e^{\sum\limits_{i=1}^q(\beta_{00_i}(+,\sigma_{0_i})+\sum\limits_{(k,l)\in
T_i}\beta_{kl}(\sigma_k,\sigma_l)+\sum\limits_{k\in
T_i}h_k(\sigma_k))}}{e^{h_0(-)}\sum\limits_{\sigma\in\Omega_0}e^{\sum\limits_{i=1}^q(\beta_{00_i}(-,\sigma_{0_i})+\sum\limits_{(k,l)\in
T_i}\beta_{kl}(\sigma_k,\sigma_l)+\sum\limits_{k\in T_i}h_k(\sigma_k))}}\\
&=e^{2B_0}\prod\limits^{q}_{i=1}\frac{\sum\limits_{\sigma\in\Omega_{T_i}}e^{\beta_{00_i}(+,\sigma_{0_i})+\sum\limits_{(k,l)\in
T_i}\beta_{kl}(\sigma_k,\sigma_l)+\sum\limits_{k\in
T_i}h_k(\sigma_k)}}{\sum\limits_{\sigma\in\Omega_{T_i}}e^{\beta_{00_i}(-,\sigma_{0_i})+\sum\limits_{(k,l)\in
T_i}\beta_{kl}(\sigma_k,\sigma_l)+\sum\limits_{k\in
T_i}h_k(\sigma_k)}}\\
\end{split}
\end{displaymath}
\begin{equation}
\begin{split}
&=e^{2B_0}
\prod\limits^{q}_{i=1}\frac{a_iZ(T_{0_i},X_i=+,\zeta_{\Lambda_i})+b_iZ(T_{0_i},X_i=-,\zeta_{\Lambda_i})}
{c_iZ(T_{0_i},X_i=+,\zeta_{\Lambda_i})+d_i Z(T_{0_i},X_i=-,\zeta_{\Lambda_i})}\\
&=e^{2B_0} \prod\limits^{q}_{i=1}\frac{a_i
R^{\zeta_{\Lambda}}_{0_i}+b_i}{c_i R^{\zeta_{\Lambda}}_{0_i}+d_i},
\end{split}
\end{equation}
where $B_0=\frac{h_0(+)-h_0(-)}{2}$, $a_i=e^{\beta_{00_i}(+,+)}$,
$b_i=e^{\beta_{00_i}(+,-)}$, $c_i=e^{\beta_{00_i}(-,+)}$,
$d_i=e^{\beta_{00_i}(-,-)}$ . Now checking the base case $t=1$ where
$R^{\zeta_{\Lambda}}_{0_i}, R^{\eta_{\Lambda}}_{0_i} \in
[0,+\infty]$, by the monotonicity of $\frac{a_i
R^{\zeta_{\Lambda}}_{0_i}+b_i}{c_i R^{\zeta_{\Lambda}}_{0_i}+d_i}$
and $\frac{a_i R^{\eta_{\Lambda}}_{0_i}+b_i}{c_i
R^{\eta_{\Lambda}}_{0_i}+d_i}$,
\begin{displaymath}
\begin{split}
\max(\frac{R^{\zeta_{\Lambda}}_0}{R^{\eta_{\Lambda}}_0}
,\frac{R^{\eta_{\Lambda}}_0}{R^{\zeta_{\Lambda}}_0})&\leq
\prod\limits^{q}_{i=1}\max(\frac{a_id_i}{b_ic_i},\frac{b_ic_i}{a_id_i})\leq
e^{4qJ}.
\end{split}
\end{displaymath}
Hence, (1) holds when $t=1$. By induction, assume that (1) holds for
$t-1$, we will show that it holds for $t$. Let
$s_i=|S(T_{0_i},0_i,t-1)|$, $i=1,2,\cdots,q$, repeating above
recursive procedure, then
\begin{displaymath}
\begin{split}
\max(\frac{R^{\zeta_{\Lambda}}_0}{R^{\eta_{\Lambda}}_0}
,\frac{R^{\eta_{\Lambda}}_0}{R^{\zeta_{\Lambda}}_0})&\leq\prod\limits^{q}_{i=1}\max(\frac{\frac{a_i
R^{\zeta_{\Lambda}}_{0_i}+b_i}{c_i
R^{\zeta_{\Lambda}}_{0_i}+d_i}}{\frac{a_i
R^{\eta_{\Lambda}}_{0_i}+b_i}{c_i
R^{\eta_{\Lambda}}_{0_i}+d_i}},\frac{\frac{a_i
R^{\eta_{\Lambda}}_{0_i}+b_i}{c_i
R^{\eta_{\Lambda}}_{0_i}+d_i}}{\frac{a_i
R^{\zeta_{\Lambda}}_{0_i}+b_i}{c_i
R^{\zeta_{\Lambda}}_{0_i}+d_i}})\\
&\leq\prod\limits^{q}_{i=1}\max(\frac{R^{\zeta_{\Lambda}}_{0_i}}
{R^{\eta_{\Lambda}}_{0_i}},\frac{R^{\eta_{\Lambda}}_{0_i}}
{R^{\zeta_{\Lambda}}_{0_i}})^{|\frac{\sqrt{a_id_i}-\sqrt{b_ic_i}}
{\sqrt{a_id_i}+\sqrt{b_ic_i}}|}\\
&\leq\prod\limits^{q}_{i=1}\max(\frac{R^{\zeta_{\Lambda}}_{0_i}}
{R^{\eta_{\Lambda}}_{0_i}},\frac{R^{\eta_{\Lambda}}_{0_i}}
{R^{\zeta_{\Lambda}}_{0_i}})^{\tanh J},
\end{split}
\end{displaymath}
where the second inequality comes from Lemma 1. According to the
hypothesis of induction $\max(\frac{R^{\zeta_{\Lambda}}_{0_i}}
{R^{\eta_{\Lambda}}_{0_i}},\frac{R^{\eta_{\Lambda}}_{0_i}}
{R^{\zeta_{\Lambda}}_{0_i}}) \leq \exp(4Js_i(\tanh{J})^{t-2})$, it's
sufficient to show
\begin{displaymath}
\begin{split}
\max(\frac{R^{\zeta_{\Lambda}}_0}{R^{\eta_{\Lambda}}_0}
,\frac{R^{\eta_{\Lambda}}_0}{R^{\zeta_{\Lambda}}_0})&\leq\prod\limits^{q}_{i=1}\exp(4Js_i(\tanh{J})^{t-1})\\
&=\exp(4Js(\tanh{J})^{t-1}),
\end{split}
\end{displaymath}
where the last equation follows by $\sum_{i=1}^qs_i=s$. This
completes the proof. \ \ \ $\Box$\\

To generalize the strong correlation decay property on trees to the
general graphs, we need to utilize the remarkable property of the
self-avoiding tree, which is implicitly stated in \cite{We06} and explicitly stated in \cite{JS06}.\\
\textbf{Lemma 4}\ (\cite{JS06}).\emph{ For two-state spin systems on
$G=(V,E)$, for any configuration $\sigma_{\Lambda}$, $\Lambda\subset
V$ and any vertex $v\in V$, then
\begin{displaymath}
P_G(X_v=+|\sigma_{\Lambda})=P_{T_{saw(v)}}(X_v=+|\sigma_{\Lambda}).
\end{displaymath}}

With Lemma 3 and 4, it is enough to prove
Theorem 1.\\\\
\textbf{Proof of Theorem 1.}  Since the maximum degree of
$T_{saw(i)}$ is also bounded by $d$, obviously, $s=|S(T_{saw(i)}, i,
t)|\leq d(d-1)^{t-1}$. According to Lemma 3 and 4,
Theorem 1 is proved. \ \ \ \ $\Box$\\

\noindent\emph{Remark:} From the proof of Theorem 1, by a similar
argument, we can get
\begin{displaymath}
|\log P_G(X_v=-|\sigma_{\Lambda})-\log
P_G(X_v=-|\eta_{\Lambda})|\leq f(t),\end{displaymath} where
$f(t)=4Jd((d-1)\tanh J)^{t-1}$.

As one of the corollaries of strong correlation decay property, we
prove there is unique Gibbs measure on an infinite bounded degree
graph (an infinite graph with maximum degree over all the degree of
its vertices $\leq d$). This generalizes original Dobrushion's
condition $$d\tanh J<1\ to\ (d-1)\tanh J<1$$ for uniqueness
of Gibbs measure of  Ising models\cite{Ge88}.\\\\
\textbf{Theorem 2.} Let $\mathbf{G}$ be an infinite graph and
$\Delta(\mathbf{G})=\sup\limits_{i\in\mathbf{G}}\{\delta_i\}$.
Assume there exists a constant $d$ such that $\Delta(\mathbf{G})\leq
d$. There is a two-state spin systems on each sub finite graph $G$
of $\mathbf{G}$ which is defined by Definition 1. If $(d-1)\tanh
J<1$, then the Gibbs measure on $\mathbf{G}$
corresponding to the two-state spin systems on sub finite graphs is unique.\\\\
\textbf{Proof.} For any given finite sub graph $G=(V,E)$ of
$\mathbf{G}$, let $\partial G =\{i:(i,j)\in\mathbf{G},i\notin G,
j\in G\}$ and $G=G_0$. Suppose there is a sequence of finite sub
graphs $G_{n-1}\subset G_n$ of $\mathbf{G}$, $n=1,2,\cdots,\infty$
such that $d_n=d(G, \partial G_n)$ goes to infinity as $n\rightarrow
\infty$. Let $G'_n=G_n\cup\partial G_n$ and denote $\zeta_n$,
$\eta_n$ any two configurations on $\partial G_n$, for
$n=1,2,\cdots,\infty$. Let $\sigma_{\Lambda}$ be any configuration
on $\Lambda$, $\Lambda\subset V$. By Proposition 2.2 in \cite{We05},
the Gibbs measure is unique on $\mathbf{G}$ if
\begin{displaymath}
\lim\limits_{n\rightarrow
\infty}\sup\limits_{\zeta_n,\eta_n}\max\limits_{\Lambda\subset V,
\sigma_{\Lambda}}|P_{G'_n}(X_{\Lambda}=\sigma_{\Lambda}|\zeta_n)-P_{G'_n}(X_{\Lambda}=\sigma_{\Lambda}|\eta_n)|=0
\end{displaymath}
holds. Let $\Lambda=\{1,2,\cdots,m\}$, where $m=|\Lambda|$. Set
$\alpha_n(1)=P_{G'_n}(X_{1}=\sigma_{1}|\zeta_n)$ and
$\beta_n(1)=P_{G'_n}(X_{1}=\sigma_{1}|\eta_n)$. Let
$\alpha_n(i)=P_{G'_n}(X_{i}=\sigma_{i}|\zeta_n, \sigma_{j},1\leq
j\leq i-1)$ and $\beta_n(i)=P_{G'_n}(X_{i}=\sigma_{i}|\eta_n,
\sigma_{j},1\leq j\leq i-1)$, $i=2,3,\cdots,n$.  The telescoping
trick gives
$$P_{G'_n}(X_{\Lambda}=\sigma_{\Lambda}|\zeta_n)=\prod\limits^m_{i=1}\alpha_n(i)$$
and
$P_{G'_n}(X_{\Lambda}=\sigma_{\Lambda}|\eta_n)=\prod\limits^m_{i=1}\beta_n(i)$.
By Theorem 1 and above remark, we know,  for each $i\in\Lambda$,
\begin{displaymath}
|\log(\frac{\alpha(i)}{\beta(i)})|\leq f(d_n),
\end{displaymath}
where $f(t)$ is the decay function $f(t)=4Jd((d-1)\tanh J)^{t-1}$.
Then $$\log(\prod\limits^m_{i=1}\frac{\alpha(i)}{\beta(i)})\leq
mf(d_n),$$ where $m=|\Lambda|$. Hence
\begin{displaymath}
\sup\limits_{\zeta_n,\eta_n}\max\limits_{\Lambda\subset V,
\sigma_{\Lambda}}\log(\frac{P_{G'_n}(X_{\Lambda}=\sigma_{\Lambda}|\zeta_n)}{P_{G'_n}(X_{\Lambda}=\sigma_{\Lambda}|\eta_n)})\leq
|V|f(d_n).
\end{displaymath}
Therefore, if $(d-1)\tanh J<1$, then
\begin{displaymath}
\lim_{n\rightarrow
\infty}\sup\limits_{\zeta_n,\eta_n}\max\limits_{\Lambda\subset V,
\sigma_{\Lambda}}\log(\frac{P_{G'_n}(X_{\Lambda}=\sigma_{\Lambda}|\zeta_n)}{P_{G'_n}(X_{\Lambda}=\sigma_{\Lambda}|\eta_n)})=0.\
\end{displaymath}
This completes the proof. \ \ \ $\Box$\\\\
\textbf{\large{4. Approximating the Partition Function}}

In the proof of Lemma 3, the calculation of the marginal probability
of the root yields a local recursive procedure.  If the tree is
truncated at height $t$, and using the recursive formula (2) to
compute the marginal probability at the root, it is easy to see that
the complexity of this procedure is linear with the number of
vertices of the truncated tree. Let $\Phi_1$ denote the whole state
space, which means $P_G(X_1=+|\Phi_1)=P_G(X_1=+)$, and
$\Phi_j=\{X_i=+, 1\leq i\leq j-1 \}$, $2\leq j\leq n+1$. Let
$\widehat{p_j}$ an estimator of conditional marginal probability
$p_j=P_G(X_j=+| \Phi_j)$, $j=1,2,\cdots,n$. The algorithm to compute
the partition function is proposed as follows.\\

\noindent\textbf{Algorithm for Partition Function $Z(G$)}\\
\textbf{Input:} $G$, a graph with vertices $\{1,\cdots,n\}$, the two-state spin systems on $G$, $\epsilon>0$, precision;\\
\textbf{Output:} $\widehat{Z(G)}$, the estimator of partition
function $Z(G)$.\\
\textbf{begin}

\textbf{For} $j$ from $1$ to $n$ \textbf{do}

\hskip 0.5cm  step 1. Set
$t_j=\frac{\log(4nJd\epsilon^{-1})}{\log((d-1)tanhJ)^{-1}}+1$,

\hskip 0.5cm  step 2. Take the vertex $j$ as root and generate the
truncated subtree $T'_{saw(j)}$ with height $t_j$

\hskip 1.7cm under the condition $\Phi_j$,

\hskip 0.5cm step 3. Set initial values be $0's$ for all the
vertices of $T'_{saw(j)}$ at height $t_j$ ,

\hskip 0.5cm step 4. Computing $\widehat{p_j}$ through $T'_{saw(j)}$
by recursive formula (2).

\textbf{End For}

Compute $\widehat{Z(G)}=Z(G,\Phi_{n+1})\prod\limits^{n}_{i=1} \widehat{p_i}^{-1}$.\\
\textbf{end}\\\\
\textbf{Theorem 3.} \emph{Let $G=(V,E)$ be a graph with vertex set
$V=\{1,2,\cdots,n\}$ and edge set $E$.
There exists a positive number $d>0$ such that $\Delta(G)\leq d$. If
$J<J_d$,  then the above algorithm provides an
FPTAS for partition function of the two-state spin systems on $G$.}\\\\
\textbf{Proof.}  According to the results of Theorem 1, there is a
function  $f(t_j)\leq \frac{\epsilon}{n}$, such that
$$e^{-\frac{\epsilon}{n}}\leq \frac{p_j}{\widehat{p_j}}\leq
e^{\frac{\epsilon}{n}},$$ under the condition that
$$t_j=\frac{\log(4nJd\epsilon^{-1})}{\log((d-1)tanhJ)^{-1}}+1=O(\log
n+\log(\epsilon^{-1})).$$ Since
$p_j=\frac{Z(G,\Phi_{j+1})}{Z(G,\Phi_{j})}$, $Z(G)$ can be expressed
as the product $Z(G)=Z(G,\Phi_{n+1})\prod\limits^{n}_{i=1}p_i^{-1}$.
Hence,
\begin{displaymath}
e^{-\epsilon}\leq
e^{(-\frac{\epsilon}{n})n}\leq\prod\limits^{n}_{i=1}
\frac{\widehat{p_i}^{-1}}{p_i^{-1}}=\frac{\widehat{Z(G)}} {Z(G)}\leq
e^{(\frac{\epsilon}{n})n}\leq e^{\epsilon}.
\end{displaymath}
The complexity of the algorithm for each $j$ in the loop is
$$O(|V(T_{saw(j)},j,t_j)|)=O((d-1)^{t_j}).$$ Thus, the total
complexity of the algorithm is
\begin{displaymath}
\sum\limits_{j}O((d-1)^{O(\log
n+\log(\epsilon^{-1}))})=nO((d-1)^{O(\log
n+\log(\epsilon^{-1}))})=O(n^{O(1)}+n(\epsilon^{-1})^{O(1)}),
\end{displaymath}
which completes the proof. \ \ \ \ $\Box$\\\\
\textbf{\large{5. Conclusions and Discussions}}

We show that the Gibbs distribution of two-state spin systems on a
bounded degree graph $G=(V,E)$ with maximum  degree $d$ exhibits
strong correlation decay when $J<J_d$. By the strong correlation
decay property and the self-avoiding tree technique, we prove the
uniqueness of Gibbs measure on an infinite bounded degree graph,
which generalizes original Dobrushion's condition $d\tanh(J)<1$ to
$(d-1)\tanh(J)<1$ for uniqueness of Gibbs measure of
(anti)ferromagnetic Ising models. Since  $J_d$ is the critical point
for uniqueness of Gibbs measure on an infinite $d$ regular tree of
Ising model. This implies that the condition for inverse temperature
is tight when restricting it on Ising model.

It is not difficult to apply our results to the sparse on average
graphs \cite{MS08} and Erd$\ddot{o}$s-R$\dot{e}$nyi random graph
$G(n,d/n)$, where each edge is chosen independently with probability
$d/n$ \cite{MS08}. We also present an FPTAS for partition functions
of two-state spin systems on the bounded degree graphs. An
interesting investigation could be whether the condition is sharp
for the general two-state spin systems.


\end{document}